\documentclass{svjour3}
\smartqed  % flush right qed marks, e.g. at end of proof
\usepackage{amssymb,amsmath}
\usepackage{graphicx}
\newcommand\rf[1]{(\ref{eq:#1})}
\newcommand\lab[1]{\label{eq:#1}}
\newcommand\nonu{\nonumber}
\newcommand\br{\begin{eqnarray}}
\newcommand\er{\end{eqnarray}}
\newcommand\be{\begin{equation}}
\newcommand\ee{\end{equation}}

\newcommand\foot[1]{\footnotemark\footnotetext{#1}}
\newcommand\lb{\lbrack}
\newcommand\rb{\rbrack}

\renewcommand\({\left(}
\renewcommand\){\right)}
\newcommand\bc{\begin{center}}
\newcommand\ec{\end{center}}

\newcommand\partder[2]{\frac{{\partial {#1}}}{{\partial {#2}}}}
\renewcommand\a{\alpha}

\renewcommand\d{\delta}

\newcommand\eps{\epsilon}
\newcommand\vareps{\varepsilon}

\newcommand\G{\Gamma}

\newcommand\h{\frac{1}{2}}
\renewcommand\k{\kappa}
\renewcommand\l{\lambda}

\newcommand\m{\mu}
\newcommand\n{\nu}
\newcommand\om{\omega}
\renewcommand\O{\Omega}

\newcommand\vp{\varphi}
\renewcommand\P{\Phi}
\newcommand\pa{\partial}

\newcommand\pr{\prime}

\renewcommand\th{\theta}

\newcommand\cP{{\mathcal P}}

\newcommand{\ct}[1]{\cite{#1}}
\newcommand{\bib}[1]{\bibitem{#1}}

\newcommand\PRL[3]{\textsl{Phys. Rev. Lett.} \textbf{#1} (#2) #3}

\newcommand\PRD[3]{\textsl{Phys. Rev.} \textbf{D#1} (#2) #3}

\newcommand\PLB[3]{\textsl{Phys. Lett.} \textbf{#1B} (#2) #3}
\newcommand\CQG[3]{\textsl{Class. Quantum Grav.} \textbf{#1} (#2) #3}

\newcommand\PR[3]{\textsl{Phys. Reports} \textbf{#1} (#2) #3}

\newcommand\IJMPD[3]{\textsl{Int. J. Mod. Phys.} \textbf{D#1} (#2) #3}

\newcommand\MPLA[3]{\textsl{Mod. Phys. Lett.} \textbf{A#1} (#2) #3}

\newcommand\vpdot{\stackrel{.}{\varphi}}
\newcommand\vpddot{\stackrel{..}{\varphi}}

\newcommand\adot{\stackrel{.}{a}}
\newcommand\addot{\stackrel{..}{a}}
\newcommand\Hdot{\stackrel{.}{H}}
\begin{document}

\title{Emergent Cosmology, Inflation and Dark Energy}
% from Spontaneous Breaking of Scale Invariance}
% from Gravity-Matter Model with Two Non-Riemannian Spacetime Volume Forms}

% \titlerunning{Emergent Cosmology from Spontaneous Breaking of Scale Invariance}
% \titlerunning{Emergent Cosmology from Non-Riemannian Spacetime Volume Forms}
\titlerunning{Emergent Cosmology, Inflation and Dark Energy}

% \author{Eduardo Guendelman \and Alexander Kaganovich \and Emil Nissimov \and 
% Svetlana Pacheva}
\author{Eduardo Guendelman \\ 
Ram\'on Herrera \and Pedro Labrana \\ 
Emil Nissimov \and Svetlana Pacheva}

\authorrunning{E. Guendelman, R. Herrera, P. Labrana, E. Nissimov, S.Pacheva}

% \institute{E. Guendelman, A. Kaganovich
\institute{Eduardo Guendelman
\at Department of Physics, Ben-Gurion University of the Negev, Beer-Sheva, Israel \\
\email{guendel@bgu.ac.il, alexk@bgu.ac.il}
\and
Ram\'on Herrera,
\at Instituto de F\'{\i}sica, Pontificia Universidad Cat\'{o}lica de
Valpara\'{\i}so,  Avenida Brasil 2950, Casilla 4059,
Valpara\'{\i}so, Chile \\
\email{ramon.herrera@ucv.cl}
\and
Pedro Labra\~{n}a,
\at Departamento de F\'{\i}sica, Facultad de Ciencias, 
Universidad del B\'{\i}o-B\'{\i}o, Casilla 5-C, Concepci\'{o}n, Chile \\
\email{plabrana@ubiobio.cl}
\and
Emil Nissimov, Svetlana Pacheva
\at Institute for Nuclear Research and Nuclear Energy,
Bulgarian Academy of Sciences, Sofia, Bulgaria  \\
\email{nissimov@inrne.bas.bg, svetlana@inrne.bas.bg}
}
\date{Received: date / Accepted: date}

\maketitle

\begin{abstract}
A new class of gravity-matter models defined in terms of two independent 
non-Riemannian volume forms (alternative generally covariant integration measure 
densities) on the space-time manifold are studied in some detail. These models
involve an additional $R^2$ (square of the scalar curvature) term as well as 
scalar matter field potentials of appropriate form so that the pertinent
action is invariant under global Weyl-scale symmetry. Scale invariance is
spontaneously broken upon integration of the equations of motion 
%%%%%  ABSTRACT CHANGE - BEGIN %%%%%
for the auxiliary volume-form degrees of freedom.
%%%%%  ABSTRACT CHANGE - END %%%%%
After performing transition to the physical Einstein frame we obtain: (i) An
effective potential for the scalar field with two flat regions which allows for 
a unified description of both early universe inflation as well as of present dark
energy epoch; (ii) For a definite parameter range the model possesses a
non-singular ``emergent universe'' solution which describes an initial phase
of evolution that precedes the inflationary phase; 
%%%%%%%%%%%%%%% ADDITION  PLANCK DATA - BEGIN  %%%%%%%%%%%%%%%%%
(iii) For a reasonable choice of the parameters the present model conforms to the
Planck Collaboration data.
%%%%%%%%%%%%%%% ADDITION  PLANCK DATA - END  %%%%%%%%%%%%%%%%%

\keywords{modified gravity theories, non-Riemannian volume forms, 
global Weyl-scale symmetry spontaneous breakdown, flat regions of scalar potential,
non-singular origin of the universe}

\PACS{04.50.Kd, % Modified theories of gravity
11.30.Qc, % Spontaneous and radiative symmetry breaking
98.80.Bp, % Origin and formation of the Universe
95.36.+x % Dark energy
}

\end{abstract}

%%%%%%%%%%%%%%%%%%%%%%%%%%%%%%%%%%%%%%%%%%%%%%%%%%%%%%%%%%%%%
%%%%%%%%%%%%%%%%%%%%%%%%%%%%%%%%%%%%%%%%%%%%%%%%%%%%%%%%%%%%%
\section{Introduction}
\label{intro}

Modern cosmology has been formulated in an attractive framework where many aspects of
the observable universe can be incorporated. In this ``standard cosmological''
framework, the early universe (cf. the books \ct{early-univ} and references therein)
starts with a period of exponential expansion called ``inflation''. 
In the inflationary period also primordial density perturbations are
generated (Ref.\ct{primordial} and references therein). The ``inflation'' 
is followed by particle creation, where the observed matter and radiation were generated
\ct{early-univ}, and finally the evolution arrives to a present phase of slowly 
accelerating universe \ct{accel-exp,accel-exp-2}.
In this standard model, however, at least two fundamental questions remain 
unanswered:  

\begin{itemize}
\item % 1)
The early inflation, although solving many cosmological puzzles,
like the horizon and  flatness problems, cannot address the initial singularity 
problem; 
\item % 2) 
There is no explanation for the existence of two periods of 
exponential expansion with such wildly different scales -- the inflationary 
phase and the present phase of slowly accelerated expansion of the universe.
\end{itemize}

The best known mechanism for generating a period of accelerated expansion 
is through the presence of some vacuum energy. In the context of a 
scalar field theory, vacuum energy density appears naturally when the scalar
field acquires an effective potential $U_{\rm eff}$ which has flat regions so 
that the scalar field can ``slowly roll'' \ct{slow-roll,slow-roll-param} and its 
kinetic energy can be neglected resulting in an energy-momentum tensor 
%% CORRECTED SIGN
$T_{\m\n} \simeq - g_{\m\n} U_{\rm eff}$.

%%%%%%%%%%%%%%%  BEGIN REVISE %%%%%%%%%%%%%%%%%%%%%%%%%
The possibility of continuously connecting an inflationary phase to a slowly 
accelerating universe through the evolution of a single scalar field -- the
{\em quintessential inflation scenario} -- has been first studied in 
Ref.\ct{peebles-vilenkin}. Also, % carefully constructed
$F(R)$ models can yield 
both an early time inflationary epoch and a late time de Sitter phase with 
vastly different values of effective vacuum energies \ct{starobinsky-2}.
For a recent proposal of a quintessential inflation mechanism based on 
the k-essence \ct{k-essence} framework, see Ref.\ct{saitou-nojiri}. For
another recent approach to quintessential inflation based on the 
``variable gravity'' model \ct{wetterich} and for extensive list of references 
to earlier work on the topic, % on quintessential inflation, \
see Ref.\ct{murzakulov-etal}.
%%%%%%%%%%%%%%%  END REVISE %%%%%%%%%%%%%%%%%%%%%%%%%

In the present paper we will study % how 
a unified scenario where both an inflation 
and a slowly accelerated phase for the universe can appear naturally from the 
existence of two flat regions in the effective scalar field potential which
we derive systematically from a Lagrangian action principle. 
Namely, we start with a new kind of globally Weyl-scale invariant gravity-matter 
action within the first-order (Palatini) approach formulated in terms of two 
different non-Riemannian volume forms (integration measures) \ct{quintess}.
In this new theory there is a single scalar field with kinetic terms coupled to 
both non-Riemannian measures, and in addition to the scalar curvature 
term $R$ also an $R^2$ term is included (which is similarly allowed by global 
Weyl-scale invariance). Scale invariance is spontaneously broken upon solving part 
of the corresponding equations of motion due to the appearance of two 
arbitrary dimensionful integration constants. We find in the physical
Einstein frame an effective k-essence \ct{k-essence} type of theory,
where the effective scalar field potential has {\em two flat regions}
corresponding to the two accelerating phases of the universe -- the
inflationary early universe and the present late universe.

In addition, within the flat region corresponding to the early universe 
we also obtain another phase that precedes the inflation and provides for a 
non-singular origin of the universe. It is of an ``emergent universe'' type 
\ct{emergent-univ}, \textsl{i.e.}, the universe starts as a static Einstein universe, 
the scalar field rolls with a constant speed through a flat region and there is a 
domain in the parameter space of the theory where such non-singular solution exists 
and is stable. To this end let us recall that the concept of ``emergent universe'' 
solves one of the principal puzzles in cosmology -- the problem of initial 
singularity \ct{singular-univ} %,\cite{HE-book}, 
including avoiding the singularity theorems for scalar field-driven inflationary
cosmology \ct{Borde-Vilenkin-PRL}. 

Let us briefly recall the origin of current approach. The main idea comes from 
Refs.\ct{TMT-orig-1}-\ct{TMT-orig-3} (see also 
Refs.\ct{TMT-recent-1-a}-\ct{TMT-recent-2}),
where some of us have proposed a new class of gravity-matter theories based on the 
idea that the action integral may contain a new metric-independent generally-covariant 
integration measure density, \textsl{i.e.}, an alternative non-Riemannian volume form 
on the space-time manifold defined in terms of an auxiliary antisymmetric gauge
field of maximal rank. The originally proposed modified-measure gravity-matter theories
\ct{TMT-orig-1}-\ct{TMT-recent-2} contained two terms in the pertinent Lagrangian action
-- one with a non-Riemannian integration measure and a second one with the
standard Riemannian integration measure (in terms of the square-root of the
determinant of the Riemannian space-time metric). An important feature was the
requirement for global Weyl-scale invariance which subsequently underwent
dynamical spontaneous breaking \ct{TMT-orig-1}. The second action term
with the standard Riemannian integration measure  might also contain a
Weyl-scale symmetry preserving $R^2$-term \ct{TMT-orig-3}.

The latter formalism yields various new interesting results
in all types of known generally covariant theories:

%%%%%%%%%%%%%%% BEGIN REVISE %%%%%%%%%%%%%%%%%%%%%%%%%
\begin{itemize}
\item
(i) $D=4$-dimensional models of gravity and matter fields containing 
the new measure of integration appear to be promising candidates for resolution 
of the dark energy and dark matter problems, the fifth force problem, 
and a natural mechanism for spontaneous breakdown of global Weyl-scale symmetry
\ct{TMT-orig-1}-\ct{TMT-recent-2}.
\item
(ii) Study of reparametrization invariant theories of extended objects 
(strings and branes) based on employing of a modified non-Riemannian 
world-sheet/world-volume integration measure \ct{mstring} leads to dynamically 
induced variable string/brane tension and to string models of non-abelian 
confinement. 
%%%%%%%%%%%% ADD CITATION %%%%%%%%%%
Recently \ct{nishino-rajpoot} this formalism was generalized to
the case of string and brane models in curved supergravity background.
%%%%%%%%%%%%%%%%%%%%%%%%%%%%%%%%%%%%
\item
(iii) Study in Refs.\ct{susy-break} of modified supergravity models with an
alternative non-Riemannian volume form on the space-time manifold produces some
outstanding new features: 
(a) This new formalism applied to minimal $N=1$ supergravity 
naturally triggers the appearance of a dynamically generated cosmological constant
as an arbitrary integration constant, which signifies a new explicit
mechanism of spontaneous (dynamical) breaking of supersymmetry;
(b) Applying the same formalism to anti-de Sitter supergravity allows us to 
appropriately choose the above mentioned arbitrary 
integration constant so as to obtain simultaneously a very small effective
observable cosmological constant as well as a very large physical gravitino mass.
\end{itemize}
%%%%%%%%%%%%%%% END REVISE %%%%%%%%%%%%%%%%%%%%%%%%%

The plan of the present paper is as follows. In the next Section 2 we describe in
some detail the general formalism for the new class of gravity-matter
systems defined in terms of two independent non-Riemannian integration
measures. In Section 3 we describe the properties of the two flat regions in the 
Einstein-frame effective scalar potential corresponding to the evolution of
the early and late universe, respectively.
%%%%%%%%%%%%%%% ADDITION  PLANCK DATA - BEGIN  %%%%%%%%%%%%%%%%%
In Section 4 we present a numerical analysis, for a reasonable choice of the 
parameters, of the resulting ratio of tensor-to-scalar 
perturbations and show that the present model conforms to the Planck Collaboration data.
%%%%%%%%%%%%%%% ADDITION  PLANCK DATA - END  %%%%%%%%%%%%%%%%%
In Section 5 we derive a non-singular ``emergent universe'' solution
of the new gravity-matter system. 
%%%%%%%%%%%%%%% ADDITION SUPERIFLATION  - BEGIN  %%%%%%%%%%%%%%%%
In Section 6 a numerical study of the transition between the
emergent universe phase and the slow-roll inflationary phase via a short
``super-inflation'' period is given in some detail.
%%%%%%%%%%%%%%% ADDITION SUPERIFLATION  - BEGIN  %%%%%%%%%%%%%%%%
We conclude in Section 7 with some discussions. 

%%%%%%%%%%%%%%%%%%%%%%%%%%%%%%%%%%%%%%%%%%%%%%%%%%%%%%%%%%%%%
\section{Gravity-Matter Formalism With Two Independent Non-Riemannian Volume-Forms}
\label{TMMT}
We shall consider the following non-standard gravity-matter system with an action 
of the general form involving two independent non-Riemannian integration
measure densities generalizing the model studied in \ct{quintess} (for simplicity 
we will use units where the Newton constant is taken as $G_{\rm Newton} = 1/16\pi$):
\be
S = \int d^4 x\,\P_1 (A) \Bigl\lb R + L^{(1)} \Bigr\rb +  
\int d^4 x\,\P_2 (B) \Bigl\lb L^{(2)} + \eps R^2 + 
\frac{\P (H)}{\sqrt{-g}}\Bigr\rb \; .
\lab{TMMT}
\ee
Here the following notations are used:

\begin{itemize}
\item
$\P_{1}(A)$ and $\P_2 (B)$ are two independent non-Riemannian volume-forms, 
\textsl{i.e.}, generally covariant integration measure densities on the underlying
space-time manifold:
\be
\P_1 (A) = \frac{1}{3!}\vareps^{\m\n\k\l} \pa_\m A_{\n\k\l} \quad ,\quad
\P_2 (B) = \frac{1}{3!}\vareps^{\m\n\k\l} \pa_\m B_{\n\k\l} \; ,
\lab{Phi-1-2}
\ee
defined in terms of field-strengths of two auxiliary 3-index antisymmetric
tensor gauge fields\foot{In $D$ space-time dimensions one can always
represent a maximal rank antisymmetric gauge field $A_{\m_1\ldots\m_{D-1}}$
in terms of $D$ auxiliary scalar fields $\phi^i$ ($i=1,\ldots,D$) in the form:
$A_{\m_1\ldots\m_{D-1}} = \frac{1}{D}\vareps_{i i_1\ldots i_{D-1}} 
\phi^i \pa_{\m_1}\phi^{i_1}\ldots \pa_{\m_{D-1}}\phi^{i_{D-1}}$, so that its
(dual) field-strength
$\P(A) = \frac{1}{D!}\vareps_{i_1\ldots i_D} \vareps^{\m_1\ldots\m_D}
\pa_{\m_1}\phi^{i_1}\ldots \pa_{\m_D}\phi^{i_D}$.}. 
$\P_{1,2}$ take over the role of the standard Riemannian integration measure density 
$\sqrt{-g} \equiv \sqrt{-\det\Vert g_{\m\n}\Vert}$ in terms of the space-time
metric $g_{\m\n}$.
\item
$R = g^{\m\n} R_{\m\n}(\G)$ and $R_{\m\n}(\G)$ are the scalar curvature and the 
Ricci tensor in the first-order (Palatini) formalism, where the affine
connection $\G^\m_{\n\l}$ is \textsl{a priori} independent of the metric $g_{\m\n}$.
Note that in the second action term we have added a $R^2$ gravity term
(again in the Palatini form). Let us recall that $R+R^2$ gravity within the
second order formalism (which was also the first inflationary model) was originally
proposed in Ref.\ct{starobinsky}.
\item
$L^{(1,2)}$ denote two different Lagrangians of a single scalar matter field of
the form (similar to the choice in Refs.\ct{TMT-orig-1}):
\br
L^{(1)} = -\h g^{\m\n} \pa_\m \vp \pa_\n \vp - V(\vp) \quad ,\quad
V(\vp) = f_1 \exp \{-\a\vp\} \; ,
\lab{L-1} \\
L^{(2)} = -\frac{b}{2} e^{-\a\vp} g^{\m\n} \pa_\m \vp \pa_\n \vp + U(\vp) 
\quad ,\quad U(\vp) = f_2 \exp \{-2\a\vp\} \; ,
\lab{L-2}
\er
where $\a, f_1, f_2$ are dimensionful positive parameters, whereas $b$ is a
dimensionless one.
\item
$\P (H)$ indicates the dual field strength of a third auxiliary 3-index antisymmetric
tensor gauge field:
\be
\P (H) = \frac{1}{3!}\vareps^{\m\n\k\l} \pa_\m H_{\n\k\l} \; ,
\lab{Phi-H}
\ee
whose presence is crucial for non-triviality of the model. 
\end{itemize}

The scalar potentials have been chosen in such a way that the original action 
\rf{TMMT} is invariant under global Weyl-scale transformations:
\br
g_{\m\n} \to \l g_{\m\n} \;\; ,\;\; \G^\m_{\n\l} \to \G^\m_{\n\l} \;\; ,\;\; 
\vp \to \vp + \frac{1}{\a}\ln \l \;\;,
\nonu \\
A_{\m\n\k} \to \l A_{\m\n\k} \;\; ,\;\; B_{\m\n\k} \to \l^2 B_{\m\n\k}
\;\; ,\;\; H_{\m\n\k} \to H_{\m\n\k} \; .
\lab{scale-transf}
\er
For the same reason we have multiplied by an appropriate exponential factor 
the scalar kinetic term in $L^{(2)}$ and also  $R$ and $R^2$ couple to the two 
different modified measures because of the different scalings of the latter.

%%%%%%%%%%%%%%%% ADDITION 1 - BEGIN  %%%%%%%%%%%%%%%%
Let us note that the requirement about the global Weyl-scale symmetry 
\rf{scale-transf} uniquely fixes the structure of the 
non-Riemannian-measure gravity-matter action \rf{TMMT} (recall that the
gravity terms $R$ and $R^2$ are taken in the first order (Palatini) formalism).
%%%%%%%%%%%%%%%% ADDITION 1 - END  %%%%%%%%%%%%%%%%

%%%%%%%%%%%%%%%% ADDITION 3 - BEGIN  %%%%%%%%%%%%%%%%
Let us also note that the global Weyl-scale symmetry transformations defined in 
\rf{scale-transf} are {\em not} the standard Weyl-scale (or conformal) symmetry
known in ordinary conformal field theory. It is straightforward to check that 
the dimensionful parameters $\a, f_1, f_2$ present in \rf{L-1}-\rf{L-2}
do {\em not} spoil at all the symmetry given in \rf{scale-transf}. 
In particular, unlike the standard form of the Weyl-scale transformation for
the metric the transformation of the scalar field $\vp$ is not the 
canonical scale transformation known in standard conformal field theories.
In fact, as shown in the second Ref.\ct{TMT-orig-1} in the context of a
simpler than \rf{TMMT} model with only one non-Riemannian measure, upon
appropriate $\vp$-dependent conformal rescaling of the metric together with a scalar field
redefinition $\vp \to \phi \sim e^{-\vp}$, one can transform the latter
model into Zee's induced gravity model \ct{zee-induced-grav}, where its
pertinent scalar field $\phi$ transforms multiplicatively under the above
scale transformations as in standard conformal field theory.
%%%%%%%%%%%%%%%% ADDITION 3 - END  %%%%%%%%%%%%%%%%

The equations of motion resulting from the action \rf{TMMT} are as follows.
Variation of \rf{TMMT} w.r.t. affine connection $\G^\m_{\n\l}$:
\be
\int d^4\,x\,\sqrt{-g} g^{\m\n} \Bigl(\frac{\P_1}{\sqrt{-g}} +
2\eps\,\frac{\P_2}{\sqrt{-g}}\, R\Bigr) \(\nabla_\k \d\G^\k_{\m\n}
- \nabla_\m \d\G^\k_{\k\n}\) = 0 
\lab{var-G}
\ee
shows, following the analogous derivation in the Ref.\ct{TMT-orig-1}, that 
$\G^\m_{\n\l}$ becomes a Levi-Civita connection:
\be
\G^\m_{\n\l} = \G^\m_{\n\l}({\bar g}) = 
\h {\bar g}^{\m\k}\(\pa_\n {\bar g}_{\l\k} + \pa_\l {\bar g}_{\n\k} 
- \pa_\k {\bar g}_{\n\l}\) \; ,
\lab{G-eq}
\ee
w.r.t. to the Weyl-rescaled metric ${\bar g}_{\m\n}$:
\be
{\bar g}_{\m\n} = (\chi_1 + 2\eps \chi_2 R) g_{\m\n} \;\; ,\;\; 
\chi_1 \equiv \frac{\P_1 (A)}{\sqrt{-g}} \;\; ,\;\;
\chi_2 \equiv \frac{\P_2 (B)}{\sqrt{-g}} \; .
\lab{bar-g}
\ee

Variation of the action \rf{TMMT} w.r.t. auxiliary tensor gauge fields
$A_{\m\n\l}$, $B_{\m\n\l}$ and $H_{\m\n\l}$ yields the equations:
\be
\pa_\m \Bigl\lb R + L^{(1)} \Bigr\rb = 0 \quad, \quad
\pa_\m \Bigl\lb L^{(2)} + \eps R^2 + \frac{\P (H)}{\sqrt{-g}}\Bigr\rb = 0 
\quad, \quad \pa_\m \Bigl(\frac{\P_2 (B)}{\sqrt{-g}}\Bigr) = 0 \; ,
\lab{A-B-H-eqs}
\ee
whose solutions read:
\be
\frac{\P_2 (B)}{\sqrt{-g}} \equiv \chi_2 = {\rm const} \;\; ,\;\;
R + L^{(1)} = - M_1 = {\rm const} \;\; ,\;\; 
L^{(2)} + \eps R^2 + \frac{\P (H)}{\sqrt{-g}} = - M_2  = {\rm const} \; .
\lab{integr-const}
\ee
Here $M_1$ and $M_2$ are arbitrary dimensionful and $\chi_2$
arbitrary dimensionless integration constants. 

%%%%%%%%%%%%%%%% ADDITION 6 - BEGIN  %%%%%%%%%%%%%%%%
The first integration constant $\chi_2$ in \rf{integr-const} preserves
global Weyl-scale invariance \rf{scale-transf}, whereas 
the appearance of the second and third integration constants $M_1,\, M_2$
signifies {\em dynamical spontaneous breakdown} of global Weyl-scale invariance 
under \rf{scale-transf} due to the scale non-invariant solutions 
(second and third ones) in \rf{integr-const}. 

To this end let us recall that classical solutions of the whole set of equations 
of motion (not only those of the scalar field(s)) correspond in the semiclassical 
limit to ground-state expectation values of the corresponding fields. 
In the present case some of the pertinent classical solutions 
(second and third Eqs.\rf{integr-const}) contain arbitrary integration constants
$M_1,\, M_2$ whose appearance makes these solutions non-covariant w.r.t. the 
symmetry transformations \rf{scale-transf}. Thus, spontaneous symmetry
breaking of \rf{scale-transf} is not necessarily originating from some fixed
extrema of the scalar potentials. In fact, as we will see in the next
Section below, the (static) classical solutions for the scalar field defined
through extremizing the effective Einstein-frame scalar potential
(Eq.\rf{U-eff} below) belong to the two infinitely large flat regions of the 
latter (infinitely large ``valleys'' of ``ground states''), therefore, this does 
not constitute a breakdown of the shift symmetry of the scalar field
\rf{scale-transf}. Thus, it is the appearance of the arbitrary integration 
constants $M_1,\, M_2$, which triggers the spontaneous breaking of 
global Weyl-scale symmetry \rf{scale-transf}. 
%%%%%%%%%%%%%%%% ADDITION 6 - END  %%%%%%%%%%%%%%%%

Varying \rf{TMMT} w.r.t. $g_{\m\n}$ and using relations \rf{integr-const} 
we have:
\be
% \chi_1 \Bigl\lb R_{\m\n} + \partder{}{g^{\m\n}} L^{(1)}\Bigr\rb -
\chi_1 \Bigl\lb R_{\m\n} + \h\( g_{\m\n}L^{(1)} - T^{(1)}_{\m\n}\)\Bigr\rb -
\h \chi_2 \Bigl\lb T^{(2)}_{\m\n} + g_{\m\n} \(\eps R^2 + M_2\)
- 2 R\,R_{\m\n}\Bigr\rb = 0 \; ,
\lab{pre-einstein-eqs}
\ee
where $\chi_1$ and $\chi_2$ are defined in \rf{bar-g},
% and first relation \rf{integr-const}, 
and $T^{(1,2)}_{\m\n}$ are the energy-momentum tensors of the scalar
field Lagrangians with the standard definitions:
\be
T^{(1,2)}_{\m\n} = g_{\m\n} L^{(1,2)} - 2 \partder{}{g^{\m\n}} L^{(1,2)} \; .
\lab{EM-tensor}
\ee

Taking the trace of Eqs.\rf{pre-einstein-eqs} and using again second relation 
\rf{integr-const} we solve for the scale factor $\chi_1$:
\be
\chi_1 = 2 \chi_2 \frac{T^{(2)}/4 + M_2}{L^{(1)} - T^{(1)}/2 - M_1} \; ,
\lab{chi-1}
\ee
where $T^{(1,2)} = g^{\m\n} T^{(1,2)}_{\m\n}$. 

Using second relation \rf{integr-const} Eqs.\rf{pre-einstein-eqs} can be put 
in the Einstein-like form:
\br
R_{\m\n} - \h g_{\m\n}R = \h g_{\m\n}\(L^{(1)} + M_1\)
+ \frac{1}{2\O}\(T^{(1)}_{\m\n} - g_{\m\n}L^{(1)}\)
\nonu \\
+ \frac{\chi_2}{2\chi_1 \O} \Bigl\lb T^{(2)}_{\m\n} + 
g_{\m\n} \(M_2 + \eps(L^{(1)} + M_1)^2\)\Bigr\rb \; ,
\lab{einstein-like-eqs}
\er
where:
\be
\O = 1 - \frac{\chi_2}{\chi_1}\,2\eps\(L^{(1)} + M_1\) \; .
\lab{Omega-eq}
\ee
Let us note that \rf{bar-g}, upon taking into account second relation
\rf{integr-const} and \rf{Omega-eq}, can be written as:
\be
{\bar g}_{\m\n} = \chi_1\O\,g_{\m\n} \; .
\lab{bar-g-2}
\ee

Now, we can bring Eqs.\rf{einstein-like-eqs} into the standard form of Einstein 
equations for the rescaled  metric ${\bar g}_{\m\n}$ \rf{bar-g-2}, 
\textsl{i.e.}, the Einstein-frame gravity equations: 
\be
R_{\m\n}({\bar g}) - \h {\bar g}_{\m\n} R({\bar g}) = \h T^{\rm eff}_{\m\n}
\lab{eff-einstein-eqs}
\ee
with energy-momentum tensor corresponding (according to \rf{EM-tensor}):
%%%%%%%%%%%%%%%% ADDITION 4A - BEGIN  %%%%%%%%%%%%%%%%
\be
T^{\rm eff}_{\m\n} = g_{\m\n} L_{\rm eff} - 2 \partder{}{g^{\m\n}} L_{\rm eff} 
\lab{EM-tensor-eff}
\ee
%%%%%%%%%%%%%%%% ADDITION 4A - END  %%%%%%%%%%%%%%%%
to the following effective Einstein-frame scalar field Lagrangian:
\be
L_{\rm eff} = \frac{1}{\chi_1\O}\Bigl\{ L^{(1)} + M_1 +
\frac{\chi_2}{\chi_1\O}\Bigl\lb L^{(2)} + M_2 + %% CORRECTION  M_1 __> M_2
\eps (L^{(1)} + M_1)^2\Bigr\rb\Bigr\} \; .
\lab{L-eff}
\ee

In order to explicitly write $L_{\rm eff}$ in terms of the Einstein-frame
metric ${\bar g}_{\m\n}$ \rf{bar-g-2} we use the short-hand notation for the
scalar kinetic term:
\be
X \equiv - \h {\bar g}^{\m\n}\pa_\m \vp \pa_\n \vp
\lab{X-def}
\ee
and represent $L^{(1,2)}$ in the form:
\be
L^{(1)} = \chi_1\O\, X - V \quad ,\quad L^{(2)} = \chi_1\O\,b e^{-\a\vp}X + U \; ,
\lab{L-1-2-Omega}
\ee
with $V$ and $U$ as in \rf{L-1}-\rf{L-2}.

From Eqs.\rf{chi-1} and \rf{Omega-eq}, taking into account \rf{L-1-2-Omega}, 
we find:
% \br
% \frac{1}{\chi_1\O} = \frac{1}{2\chi_2} (V-M_1)
% \Bigl\lb 1+2\eps\chi_2\,X\, \Bigl( 1 + \frac{V-M_1}{2(U+M_2)}b
% e^{-\a\vp}\Bigr)\Bigr\rb % \Bigl\lb U+M_2 + \eps (V-M_1)^2\Bigr\rb^{-1}
% \nonu \\ 
% \times \Bigl\lb U+M_2 + \eps (V-M_1)^2\Bigr\rb^{-1}
% - \frac{b e^{-\a\vp}}{2(U+M_2)}\, X \; .
% \lab{chi-Omega}
% \er
\be
\frac{1}{\chi_1\O} = \frac{(V-M_1)}{2\chi_2\Bigl\lb U+M_2 + \eps (V-M_1)^2\Bigr\rb}
\,\Bigl\lb 1 - \chi_2 \Bigl(\frac{b e^{-\a\vp}}{V-M_1} - 2\eps\Bigr) X\Bigr\rb \; .
\lab{chi-Omega}
\ee
Upon substituting expression \rf{chi-Omega} into \rf{L-eff} we arrive at the
explicit form for the Einstein-frame scalar Lagrangian:
\be
L_{\rm eff} = A(\vp) X + B(\vp) X^2 - U_{\rm eff}(\vp) \; ,
\lab{L-eff-final}
\ee
where:
\br
A(\vp) \equiv 1 + \Bigl\lb \h b e^{-\a\vp} - \eps (V - M_1)\Bigr\rb
\frac{V - M_1}{U + M_2 + \eps (V - M_1)^2}
\nonu\\
= 1 + \Bigl\lb \h b e^{-\a\vp} - \eps\(f_1 e^{-\a\vp} - M_1\) \Bigr\rb
\,\frac{f_1 e^{-\a\vp} - M_1}{f_2 e^{-2\a\vp} + M_2 + \eps (f_1 e^{-\a\vp} - M_1)^2}
\; , 
\lab{A-def}
\er
and
\br
B(\vp) \equiv \chi_2 \frac{\eps\Bigl\lb U + M_2 + (V - M_1) b e^{-\a\vp}\Bigr\rb
- \frac{1}{4} b^2 e^{-2\a\vp}}{U + M_2 + \eps (V - M_1)^2}
\nonu\\
= \chi_2 \frac{\eps\Bigl\lb f_2 e^{-2\a\vp} + M_2 +
(f_1 e^{-\a\vp} - M_1)b e^{-\a\vp}\Bigr\rb -\frac{1}{4} b^2 e^{-2\a\vp}
}{f_2 e^{-2\a\vp} + M_2 + \eps (f_1 e^{-\a\vp} - M_1)^2} \; ,
\lab{B-def}
\er
whereas the effective scalar field potential reads:
\be
U_{\rm eff} (\vp) \equiv 
\frac{(V - M_1)^2}{4\chi_2 \Bigl\lb U + M_2 + \eps (V - M_1)^2\Bigr\rb}
= \frac{\(f_1 e^{-\a\vp}-M_1\)^2}{4\chi_2\,\Bigl\lb 
f_2 e^{-2\a\vp} + M_2 + \eps (f_1 e^{-\a\vp}-M_1)^2\Bigr\rb} \; ,
\lab{U-eff}
\ee
where the explicit form of $V$ and $U$ \rf{L-1}-\rf{L-2} are inserted.

Let us recall that the dimensionless integration constant $\chi_2$ is the
ratio of the original second non-Riemannian integration measure to the
standard Riemannian one \rf{bar-g}.

%%%%%%%%%%%%%%%% ADDITION 2 - BEGIN  %%%%%%%%%%%%%%%%
To conclude this Section let us note that choosing  the ``wrong'' sign of the scalar 
potential $U(\vp)$ (Eq.\rf{L-2}) in the initial non-Riemannian-measure
gravity-matter action \rf{TMMT} is necessary to end up with the right sign in the
effective scalar potential \rf{U-eff} in the physical Einstein-frame effective 
gravity-matter action \rf{L-eff-final}. On the other hand, the overall sign of the
other initial scalar potential $V(\vp)$ (Eq.\rf{L-2}) is in fact irrelevant since
changing its sign does not affect the positivity of effective scalar potential
\rf{U-eff}.
%%%%%%%%%%%%%%%% ADDITION 2 - END  %%%%%%%%%%%%%%%%

%%%%%%%%%%%%%%%% ADDITION 4B - END  %%%%%%%%%%%%%%%%
Let us also remark that the effective matter Lagrangian \rf{L-eff-final} is called
``Einstein-frame scalar Lagrangian'' in the sense that it produces the 
effective energy-momentum tensor \rf{EM-tensor-eff} entering the effective 
Einstein-frame form of the gravity equations of motion \rf{eff-einstein-eqs}
in terms of the conformally rescaled metric ${\bar g}_{\m\n}$ \rf{bar-g-2}
which have the canonical form of Einstein's gravitational equations. On the
other hand, the pertinent Einstein-frame effective scalar Lagrangian 
\rf{L-eff-final} arises in a non-canonical ``k-essence'' \ct{k-essence} type
form.
%%%%%%%%%%%%%%%% ADDITION 4B - END  %%%%%%%%%%%%%%%%

%%%%%%%%%%%%%%%%%%%%%%%%%%%%%%%%%%%%%%%%%%%%%%%%%%%%%%%%%%%%%
%%%%%%%%%%%%%%%%%%%%%%%%%%%%%%%%%%%%%%%%%%%%%%%%%%%%%%%%%%%%%
\section{Flat Regions of the Effective Scalar Potential}
\label{flat-regions}

Depending on the sign of the integration constant $M_1$ we obtain two types of 
shapes for the effective scalar potential $U_{\rm eff} (\vp)$ \rf{U-eff}
depicted on Fig.1 and Fig.2.
% %%%%%%%%%%%%%%%%%%%%%%%%%%%%% QUALITATIVE REPRESENTATION %%%%%%%%%%%%
Due to the vast difference in the scales of the pertinent parameters, whose
estimates are given below, Fig.1 and Fig.2 represent only qualitatively the 
shape of $U_{\rm eff} (\vp)$.
% %%%%%%%%%%%%%%%%%%%%%%%%%%%%%%%%%%%%%%%%%%%%%%%%%%%%%%%%%%%%%%%%%%%%%

% %%%%%%%%%%%%%%%%%%%%%%%%%%%%% Adding ``qualitative'' %%%%%%%%%%%%%%%%
\begin{figure}
\begin{center}
\includegraphics{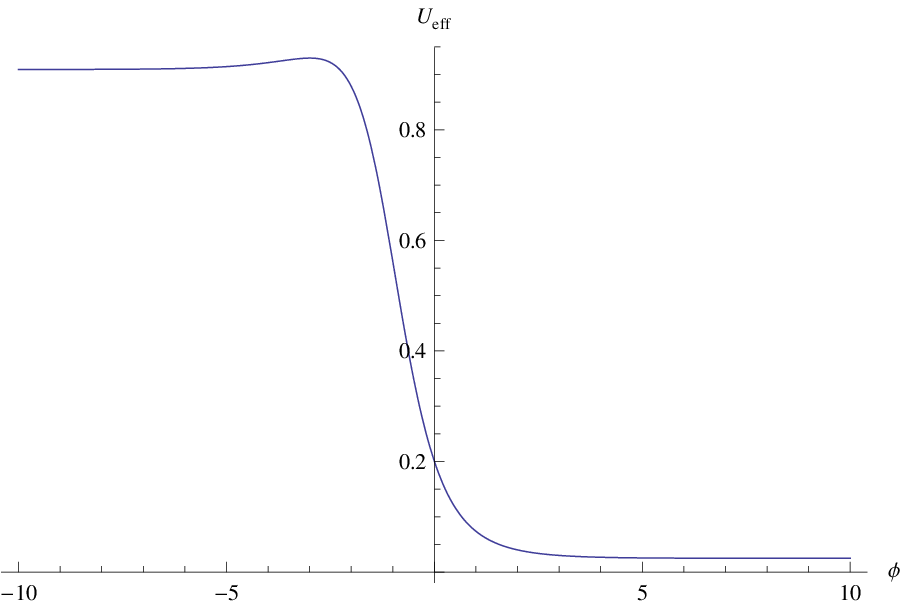}
\caption{Qualitative shape of the effective scalar potential $U_{\rm eff}$ \rf{U-eff}
as function of $\vp$ for $M_1 < 0$.}
\end{center}
\end{figure}
%%%%%%%%%%%%%%%%%%%%%%%%%%%%% Adding ``qualitative'' %%%%%%%%%%%%%%%%
\begin{figure}
\begin{center}
\includegraphics{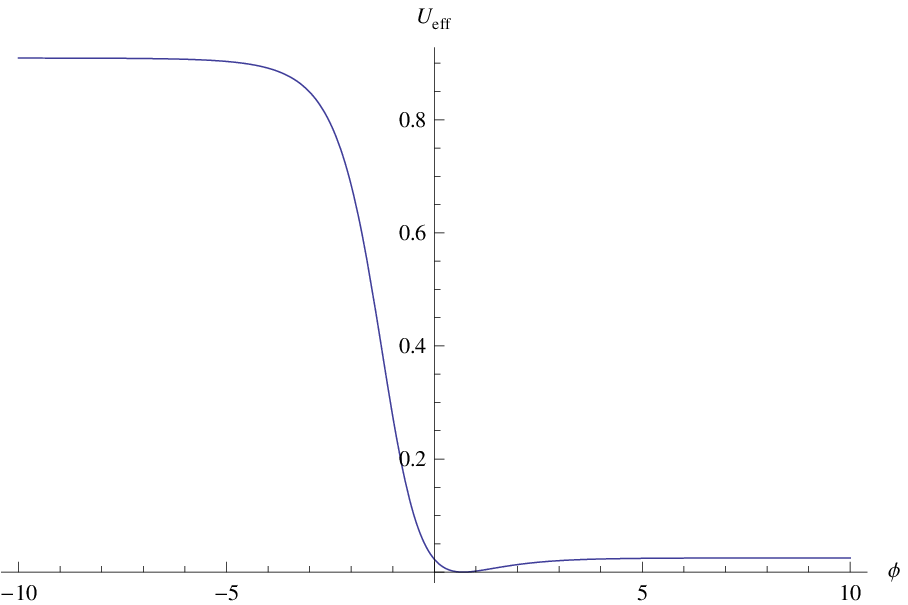}
\caption{Qualitative shape of the effective scalar potential $U_{\rm eff}$ \rf{U-eff}
as function of $\vp$ for $M_1 > 0$.}
\end{center}
\end{figure}
%%%%%%%%%%%%%%%%%%%%%%%%%%%%%

The crucial feature of $U_{\rm eff} (\vp)$ is the presence of two infinitely large
flat regions -- for large negative and large positive values of the scalar field $\vp$.
For large negative values of $\vp$ we have for the effective potential and the
coefficient functions in the Einstein-frame scalar Lagrangian 
\rf{L-eff-final}-\rf{U-eff}:
\br
U_{\rm eff}(\vp) \simeq U_{(-)} \equiv 
\frac{f_1^2/f_2}{4\chi_2 (1+\eps f_1^2/f_2)} \; ,
\lab{U-minus} \\
A(\vp) \simeq A_{(-)} \equiv \frac{1+\h b f_1/f_2}{1+\eps f^2_1/f_2} \;\; ,\;\; 
B(\vp) \simeq B_{(-)} \equiv 
- \chi_2 \frac{b^2/4f_2 - \eps (1+ b f_1/f_2)}{1+\eps f^2_1/f_2} \; .
\lab{A-B-minus}
\er
In the second flat region for large positive $\vp$:
\br
U_{\rm eff}(\vp) \simeq U_{(+)} \equiv 
\frac{M_1^2/M_2}{4\chi_2 (1+\eps M_1^2/M_2)} \; ,
\lab{U-plus} \\
A(\vp) \simeq A_{(+)} \equiv \frac{M_2}{M_2 + \eps M_1^2} \quad ,\quad
B(\vp) \simeq B_{(+)} \equiv \eps\chi_2 \frac{M_2}{M_2 + \eps M_1^2} \; .
\lab{A-B-plus}
\er

% ===============================
% MEANING OF THE TWO FLAT REGIONS
% ===============================
From the expression for $U_{\rm eff} (\vp)$ \rf{U-eff} and the figures 1 and 2 we
see that now we have an explicit realization of quintessential inflation scenario.
The flat regions \rf{U-minus}-\rf{A-B-minus} and \rf{U-plus}-\rf{A-B-plus} correspond 
to the evolution of the early and the late universe, respectively, provided we
choose the ratio of the coupling constants in the original scalar potentials 
versus the ratio of the scale-symmetry breaking integration constants to obey:
\be
\frac{f_1^2/f_2}{1+\eps f_1^2/f_2} \gg \frac{M_1^2/M_2}{1+\eps M_1^2/M_2} \, ,
\lab{early-vs-late}
\ee
which makes the vacuum energy density of the early universe $U_{(-)}$ much bigger
than that of the late universe $U_{(+)}$ (cf. \rf{U-minus}, \rf{U-plus}).
The inequality \rf{early-vs-late} is equivalent to the requirements:
\be
\frac{f_1^2}{f_2} \gg \frac{M_1^2}{M_2} \quad ,\quad |\eps| \frac{M_1^2}{M_2} \ll 1 \; .
\lab{early-vs-late-2}
\ee
In particular, % if we take the integration constant $\chi_2 \sim 1$, and 
if we choose the scales of the scale symmetry breaking integration constants
$|M_1| \sim M^4_{EW}$ and $M_2 \sim M^4_{Pl}$, where $M_{EW},\, M_{Pl}$ are
the electroweak and Plank scales, respectively, we are then naturally led to
a very small vacuum energy density $U_{(+)}\sim M_1^2/M_2$ of the order:
\be
U_{(+)}\sim M^8_{EW}/M^4_{Pl} \sim 10^{-120} M^4_{Pl} \; ,
\lab{U-plus-magnitude}
\ee
which is the right order of magnitude for the present epoche's vacuum energy density 
as already recognized in Ref.\ct{arkani-hamed}.
On the other hand, if we take the order of magnitude of the coupling constants 
in the effective potential $f_1 \sim f_2 \sim (10^{-2} M_{Pl})^4$, then
together with the above choice of order of magnitudes for $M_{1,2}$ the
inequalities \rf{early-vs-late-2} will be satisfied as well and the order of
magnitude of the vacuum energy density of the early universe $U_{(-)}$ 
\rf{U-minus} becomes:
\be
U_{(-)} \sim f_1^2/f_2 \sim 10^{-8} M_{Pl}^4 \; ,
\lab{U-minus-magnitude}
\ee
which conforms to the BICEP2 experiment \ct{bicep2} and Planck Collaboration
data \ct{Planck1,Planck2} implying the energy scale of inflation of order 
$10^{-2} M_{Pl}$.
%%%%%%%%%%%%%%%%%%%%  BICEP2 vs Planck - BEGIN  %%%%%%%%%%%%%%%%%%%%%%%%
However, let us remark at this point that, as shown in the next Section 4, the 
result for the tensor-to-scalar ratio $r$ obtained within the present model conforms 
to the data of the Planck Collaboration \ct{Planck1,Planck2} rather than 
BICEP2 \ct{bicep2}.
%%%%%%%%%%%%%%%%%%%%  BICEP2 vs Planck - END  %%%%%%%%%%%%%%%%%%%%%%%%

Let us recall that, since we are using units where $G_{\rm Newton} =
1/16\pi$, in the present case $M_{Pl}= \sqrt{1/8\pi G_{\rm Newton}} = \sqrt{2}$.

Before proceeding to the derivation of the non-singular ``emergent
universe'' solution describing an initial phase of the universe evolution
preceding the inflationary phase, let us briefly sketch how the present 
non-Riemannian-measure-modified gravity-matter theory meets the conditions
for the validity of the ``slow-roll'' approximation \ct{slow-roll} when
$\vp$ evolves on the flat region of the effective potential corresponding to the
early universe \rf{U-minus}-\rf{A-B-minus}. 

To this end let us recall the standard Friedman-Lemaitre-Robertson-Walker
space-time metric \ct{weinberg-72}:
\be
ds^2 = - dt^2 + a^2(t) \Bigl\lb \frac{dr^2}{1-K r^2}
+ r^2 (d\th^2 + \sin^2\th d\phi^2)\Bigr\rb
\lab{FLRW}
\ee
and the associated Friedman equations
(recall the presently used units $G_{\rm Newton} = 1/16\pi$):
\be
\frac{\addot}{a}= - \frac{1}{12} (\rho + 3p) \quad ,\quad
% \frac{\adot^2}{a^2} + \frac{k}{a^2} = \frac{1}{6}\rho \; ,
H^2 + \frac{K}{a^2} = \frac{1}{6}\rho \quad ,\;\; H\equiv \frac{\adot}{a} \; ,
\lab{friedman-eqs}
\ee
describing the universe' evolution. Here:
\br
\rho = \h A(\vp) \vpdot^2 + \frac{3}{4} B(\vp) \vpdot^4 + U_{\rm eff}(\vp) \; ,
\lab{rho-def} \\
p = \h A(\vp) \vpdot^2 + \frac{1}{4} B(\vp) \vpdot^4 - U_{\rm eff}(\vp)
\lab{p-def}
\er
are the energy density and pressure of the scalar field $\vp = \vp (t)$. 
Henceforth the dots indicate derivatives with respect to the time $t$.

Let us now consider the standard ``slow-roll'' parameters \ct{slow-roll-param}:
\be
\vareps \equiv - \frac{\Hdot}{H^2} \quad, \quad 
\eta \equiv -\frac{\vpddot}{H\vpdot} \; ,
\lab{SLR-def}
\ee
where $\vareps$ measures the ratio of the scalar field kinetic energy relative to
its total energy density and $\eta$ measures the ratio of the field's acceleration
relative to the ``friction'' ($\sim 3 H\vpdot$) term in the pertinent scalar
field equations of motion:
\be
\vpddot (A+3B\vpdot^2) + 3 H \vpdot (A+B\vpdot^2) + U^{\pr}_{\rm eff} 
+ \h A^{\pr}\vpdot^2 + \frac{3}{4} B^{\pr}\vpdot^4 = 0 \; ,
\lab{vp-eqs-full}
\ee
with primes indicating derivatives w.r.t. $\vp$.

In the slow-roll approximation one ignores the terms with $\vpddot$, $\vpdot^2$, 
$\vpdot^3$, $\vpdot^4$ so that the $\vp$-equation of motion \rf{vp-eqs-full}
and the second Friedman Eq.\rf{friedman-eqs} reduce to:
\be
3AH\vpdot + U^{\pr}_{\rm eff} = 0 \quad ,\quad H^2 = \frac{1}{6} U_{\rm eff} \; .
\lab{slow-roll-eqs}
\ee
The reason for ignoring the spatial curvature term $K/a^2$ in the second
Eq.\rf{slow-roll-eqs} is due to the fact that $\vp$ evolves on a flat region of 
$U_{\rm eff}$ and the Hubble parameter $H\equiv \adot\!\!/a \simeq {\rm const}$, 
so that $a(t)$ grows exponentially with time making $K/a^2$ very small.
Consistency of the slow-roll approximation implies for the slow-roll
parameters \rf{SLR-def}, taking into account \rf{slow-roll-eqs}, the following
inequalities:
\be
\vareps \simeq \frac{1}{A}\,\Bigl(\frac{U^{\pr}_{\rm eff}}{U_{\rm eff}}\Bigr)^2
\ll 1 \quad ,\quad
\eta \simeq \frac{2}{A}\,\frac{U^{\pr\pr}_{\rm eff}}{U_{\rm eff}} - \vareps
- \frac{2A^{\pr}}{A^{3/2}}\sqrt{\vareps} \;\; \to \;\; 
\frac{2}{A}\,\frac{U^{\pr\pr}_{\rm eff}}{U_{\rm eff}} \ll 1 \; .
\lab{SLR-small}
\ee

Since now $\vp$ evolves on the flat region of $U_{\rm eff}$ for large negative values
\rf{U-minus}, the Lagrangian coefficient function $A(\vp) \simeq A_{(-)}$ as
in \rf{A-B-minus} and the gradient of the effective scalar potential is:
\be
U^{\pr}_{\rm eff} \simeq 
-\frac{\a f_1 M_1 e^{\a\vp}}{2\chi_2f_2 (1+\eps f_1^2/f_2)^2} \; ,
\lab{U-prime-minus}
\ee
which yields for the slow-roll parameter $\vareps$ \rf{SLR-small}:
\be
\vareps \simeq 
\frac{4\a^2 M_1^2 e^{2\a\vp}}{f_1^2 (1+b f_1/2f_2)(1+\eps f_1^2/f_2)} \ll 1
\;\;\; {\rm for ~large ~negative} ~\vp \; .
\lab{SLR-eps}
\ee
Similarly, for the second slow-roll parameter we have:
\be
|\frac{2}{A}\,\frac{U^{\pr\pr}_{\rm eff}}{U_{\rm eff}}| \simeq
\frac{4\a^2 |M_1| e^{\a\vp}}{f_1 (1+b f_1/2f_2)} \ll 1
\;\;\; {\rm for ~large ~negative} ~\vp \; .
\lab{SLR-eta}
\ee

%%%%%%%%%%%%%%%% ADDITION 11 - BEGIN  %%%%%%%%%%%%%%%%
At this point let us remark that the non-canonical ``k-essence'' form of the
effective scalar Lagrangian \rf{L-eff-final} does not affect the condition
for smallness of the standard ``slow-roll'' parameters \rf{SLR-def}. Indeed,
the definition of the first slow-roll parameter $\vareps$ in \rf{SLR-def}
is consistent with the first Friedman equation in \rf{friedman-eqs}, where there 
is no \textsl{a priori} requirement for the energy density and the pressure to
be defined in terms of a scalar field action with a canonical kinetic term.
Similarly, the non-canonical ``k-essence'' form of the effective scalar 
Lagrangian \rf{L-eff-final} does not affect the requirement for smallness of 
the second ``slow-roll'' parameter $\eta$ in \rf{SLR-def}. In fact, the smallness 
of $\eps,\eta$ is explicitly displayed in Eqs.\rf{SLR-eps}-\rf{SLR-eta} because of the 
presence of strongly suppressing factors -- exponentials of large negative values
of the scalar field in the first flat region of the effective scalar potential
corresponding to the early universe.
%%%%%%%%%%%%%%%% ADDITION 11 - END  %%%%%%%%%%%%%%%%

The value of $\vp$ at the end of the slow-roll regime $\vp_{\rm end}$ is determined
from the condition $\vareps \simeq 1$ which through \rf{SLR-eps} yields:
\be
e^{-2\a\vp_{\rm end}} \simeq
\frac{4\a^2 M_1^2}{f_1^2 (1+b f_1/2f_2)(1+\eps f_1^2/f_2)} \; .
\lab{vp-end}
\ee
% The amount of inflation when $\vp$ evolves from some initial value $\vp_{\rm in}$ 
% to the end-point of slow-roll inflation  $\vp_{\rm end}$ is determined through the 
% expression for the
%%%%%%%%%%%%%%%% MODIFICATION  RAMON - BEGIN  %%%%%%%%%%%%%%%%
The number of {\em e-foldings} $N$ (see, \textsl{e.g.} second Ref.\ct{primordial})
between two values of cosmological times $t_{*}$ and $t_{end}$
or analogously between two different values $\vp_{*}$ and $\vp_{end}$ becomes:
\be
N = \int_{t_{*}}^{t_{\rm end}} H dt =
\int_{\vp_{*}}^{\vp_{\rm end}} \frac{H}{\vpdot} d\vp \simeq
- \int_{\vp_{*}}^{\vp_{\rm end}} \frac{3H^2 A}{U^{\pr}_{\rm eff}} d\vp \simeq
- \int_{\vp_{*}}^{\vp_{\rm end}} \frac{A U_{\rm eff}}{2 U^{\pr}_{\rm eff}} d\vp
\; ,
\lab{e-foldings}
\ee
where Eqs.\rf{slow-roll-eqs} are used. Substituting \rf{U-minus}, \rf{A-B-minus}
and \rf{U-prime-minus} into \rf{e-foldings} yields an expression for $N$ which 
together with \rf{vp-end} allows for the determination of $\vp_{*}$:
\be
N \simeq \frac{f_1 (1 + b f_1/f_2)}{4\a^2 M_1} 
\Bigl( e^{-\a\vp_{*}} - e^{-\a\vp_{\rm end}}\Bigr) \; .
\lab{vp-star-eq}
\ee
In what follows the subscript $*$ is used to indicate the epoch where the 
cosmological scale exits the horizon.
%%%%%%%%%%%%%%%% MODIFICATION  RAMON - END  %%%%%%%%%%%%%%%%

%%%%%%%%%%%%%%%%%%%%%%%%%%%%%%%%%%%%%%%%%%%%%%%%%%%%%%%%%%%%%
%%%%%%%%%%%%%%%% ADDITION  RAMON -  BEGIN  %%%%%%%%%%%%%%%%%%

\section{Perturbations}
\label{perturbations}

In the following we will describe the scalar and tensor
perturbations for our model. Following Refs.\cite{Garriga,p1}
the power spectrum of the scalar perturbation ${\cP_S}$
for a non-canonical kinetic term in the slow-roll approximation is given by:
\begin{equation}
{\cP_S} = k_1\,\frac{H^2}{c_s\,\varepsilon_1} \; ,\label{pec}
\end{equation}
where $c_s$ denotes the ``speed of sound'' and is defined as  
$c_s^2=\frac{P_{,\,X}}{P_{,\,X}+2XP_{,\,XX}}$,  and
$\varepsilon_1=X\,P_{,\,X}/(16\pi^2\,H^2)$. Here
$P(X,\varphi)$ is a function of the scalar field $\varphi$ and $X$ is the
scalar kinetic term as in \rf{X-def}.
% $X=-(1/2)\;g^{\,\mu\nu}\partial_\mu \varphi\partial_\nu \varphi$.  
The constant $k_1=(G_{Newton}/8\pi^2 )=(16\times 8\pi^3)^{-1}$,
and $P_{,\,X}$ denotes the derivative with respect $X$.
In particular, in the present case
$P(X,\vp) = L_{\rm eff} = A(\vp)\,X+B(\vp)\,X^2-U_{\rm eff}(\vp)$ \rf{L-eff-final} 
where $X=\dot{\varphi}^2/2$.

The scalar spectral index $n_s$ is given by:
\be
n_s-1=\frac{d\ln \cP_S}{d\ln k}=
-2\varepsilon_1-\varepsilon_2-\varepsilon_3 \; ,
\lab{ns}
\ee
where the parameters $\varepsilon_2$ and $\varepsilon_3$ are defined as
$\varepsilon_2=\frac{\dot{\varepsilon_1}}{\varepsilon_1\,H}$ and
$\varepsilon_3=\frac{\dot{c_s}}{c_s\,H}$, respectively \cite{Garriga,p1}.

On the other hand, it is well known that the generation of tensor perturbations during
inflation  would generate gravitational waves. The spectrum of the tensor perturbations
$\cP_T$ was calculated in Ref.\cite{Garriga} and is given by:
\begin{equation}
\cP_T=\frac{2}{3\pi^2}\,\left(\frac{2XP_{,\,X}-P}{(16\pi)^2}\right) \; ,
\label{Pt}
\end{equation}
and the tensor spectral index $n_T$ can be expressed in terms of
the parameter $\varepsilon_1$ as $ n_T=\frac{d\ln \cP_T}{d\ln k}=-2\varepsilon_1$.  
An important observational quantity is the tensor-to-scalar ratio
$r=\frac{\cP_T}{\cP_S}$ satisfying a generalized consistency relation in which 
$r=-8\,c_s\,n_T$. These observational quantities should be evaluated at 
$\varphi=\varphi_{*}$ (see Eq.\rf{vp-star-eq}).

Considering the slow-roll approximation the power spectrum of the
scalar perturbation $\cP_S$ (to leading order) from Eq.(\ref{pec}) becomes:
\begin{equation}\label{primordialpert.corr}
\cP_S \simeq\,k_1\,C_1\,e^{-2\alpha\varphi_*},
\end{equation}
where the constant $C_1$ is given by
$$
C_1=\frac{(1+\frac{bf_1}{2f_2})}{16\chi_2\alpha^2\,M_1^2}\,\frac{f_1^4}{18\,f_2}\; .
$$

From Eq.\rf{ns} the scalar spectral index $n_s$, becomes:
\be
n_s\simeq 1-\frac{2\alpha^2M_1}{f_1(1+\epsilon
f_1^2/f_2)}\,e^{\alpha\varphi_*}-\frac{4\alpha^2M_1^2}{f_1^2(1+\epsilon
f_1^2/f_2)}\left[\frac{1}{4\pi^2(1+bf_1/2f_2)}+1\right]\,e^{2\alpha\varphi_*} \; .
\lab{ns-1}
\ee

Combining Eqs.(45) and \rf{ns-1} the scalar spectral index $n_s$ can be expresses in 
terms of the number of e-foldings $N$ \rf{e-foldings} to give:
\br
n_s \simeq 1-\frac{\alpha(1+bf_1/2f_2)}{(1+\epsilon f_1^2/f_2)}[C_2+2\alpha\,N]^{-1}
\nonu \\
-\frac{(1+bf_1/2f_2)^2}{(1+\epsilon f_1^2/f_2)}
\left[\frac{1}{4\pi^2(1+bf_1/2f_2)}+1\right]\,[C_2+2\alpha\,N]^{-2} \; .
\lab{ns-2}
\er
Here we took into account the relation between $\vp_{*}$ and the number of
e-foldings $N$ \rf{vp-star-eq}, which can be written as:
\be
e^{\alpha\varphi_*}=\frac{f_1(1+bf_1/2f_2)}{2\alpha M_1}\,[C_2+2\alpha\,N]^{-1} \; ,
\lab{vp-star-2}
\ee
where the constant $C_2$ is given by:
\be
C_2=\sqrt{\frac{(1+bf_1/2f_2)}{(1+\epsilon f_1^2/f_2)}} \; .
\lab{C2-def}
\ee
%%%%%%%%%%%%%%%%%%%%%%%%%%%%%%%%%%%%%%%%%%%%%%%%%%%%%%%%%%%%%%%%
%% LAST ADDITION RAMON - 1

From Eqs.(\ref{primordialpert.corr}) and \rf{vp-star-2} we can write the parameter 
$\chi_2$ in terms of the number of e-folds $N$ and the power spectrum as:
\begin{equation}
\chi_2=\frac{k_1\,f_1^2\,[C_2+2\alpha\,N]^2}{72\,f_2(1+bf_1/2f_2)}\,
\frac{1}{\cP_S} \; .
\label{chi2}
\end{equation}

In this form, we can obtain the value of the parameter $\chi_2$
for given values $f_1$, $f_2$,$b$, $\epsilon$ and $\alpha$
parameters when the number of e-folds $N$  and the power spectrum
$\cP_S$ are given.
%%%%%%%%%%%%%%%%%%%%%%%%%%%%%%%%%%%%%%%%%%%%%%%%%%%%%%%%%%%%%%%%

From Eq.\rf{ns-2} and considering that $r=16 c_s \,\varepsilon_1$,  
the relation between the tensor-to-scalar ratio $r$ and the spectral index $n_s$ ,
\textsl{i.e.}, the consistency relation, $n_s=n_s(r)$, is given by:
\be
n_s \simeq 1-\frac{\pi\,\alpha\,C_2}{\sqrt{2}}\,r^{1/2}-\frac{\pi^2\,(1+bf_1/2f_2)}{2}\,
\left[\frac{1}{4\pi^2(1+bf_1/2f_2)}+1\right]\,r \, .
\lab{nsr}
\ee
Here we note that working to leading order the consistency relation $n_s=n_s(r)$, 
becomes independent of the integration constants $M_1$ and $\chi_2$ \rf{integr-const}.

\begin{figure}[th]
{\hspace{-6.0cm}\includegraphics[width=7.0in,angle=0,clip=true]{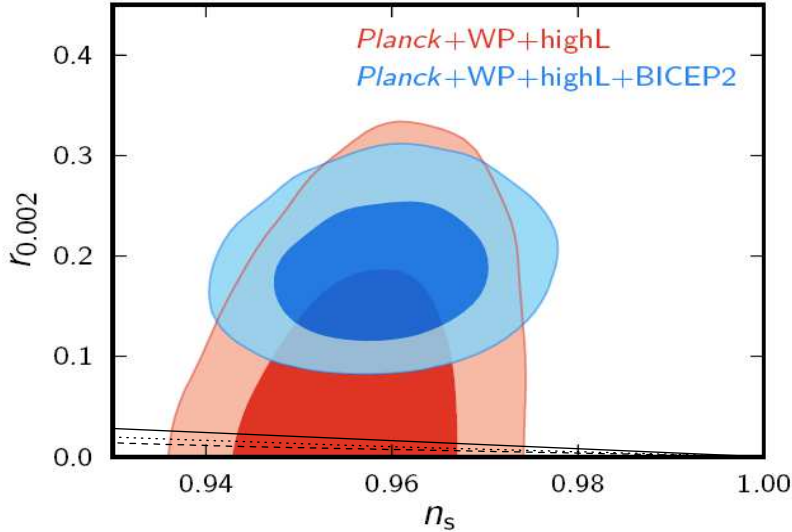}}
{\vspace{-3.5cm}\caption{Evolution of the tensor-scalar ratio $r$
versus the scalar spectrum index $n_s$, for three different value
of the parameter $\alpha$. The dashed, dotted, and solid lines are
for the values of $\alpha=0.2$, $\alpha=10^{-2}$ and
$\alpha=10^{-20}$, respectively. Also, in this plot we have taken
the values $f_1=2\times 10^{-8}$, $f_2=10^{-8}$, $\epsilon=1$,
$b=-0.52$  and $M_p=\sqrt{2}$.
\label{fig3}}}
\end{figure}

In Fig.\ref{fig3} we show the evolution of  the tensor-to-scalar
ratio $r$  w.r.t. the scalar spectral index $n_s$ for  three different
values of the parameter $\alpha$. Here we show the two-dimensional marginalized
constraints, at 68$\%$ and 95$\%$ levels of confidence, for the
tensor-to-scalar ratio $r$ and the  spectral index $n_s$ from BICEP2
experiment  in connection with Planck + WP + highL
\cite{bicep2}. In order to write down values that relate the ratio $r$ and the 
spectral index $n_s$ we considered the consistency relation $n_s=n_s(r)$ given 
by Eq.\rf{nsr}. Also, we have used the values $f_1=2\times 10^{-8}$, $f_2=10^{-8}$,
$\epsilon=1$, $b=-0.52$ and $M_p=\sqrt{2}$.

From the plot in Fig.\ref{fig3} we note that the tensor-to-scalar ratio $r\sim 0$, and 
our model is disproved from BICEP2, since according to the latter the ratio 
$r=0.2_{-0.05}^{+0.07}$ with the ratio $r=0$ disproved at 7.0$\sigma$. 
Nevertheless, the result for tensor-to-scalar ratio has become less clear when serious 
criticisms of BICEP2 appeared in the literature. 
In particular, the Planck Collaboration has issued the data % relating 
about the polarized dust
emission through an analysis of the polarized thermal emissions from diffuse Galactic dust,
which suggest that BICEP2 data of the gravitational wave result could be due to the dust 
contamination \cite{Planck1}. Thereby, a detailed analysis of Planck and BICEP2 data 
would be required for a definitive answer. In this form, previous CMB observations
from the Planck satellite and other CMB experiments obtained only an upper limit for 
the tensor-to-scalar ratio, in which $r <$ 0.11 (at 95$\%$ confidence level)
\cite{Planck2}. Therefore, we find that the value $\alpha\sim 0$ is well  supported 
by the confidence levels from Planck data. In particular, % for 
the value $\alpha=10^{-20}$
corresponds to $r\mid_{n_s=0.96}\simeq0.017$. Also, we  note that when we increase the
value of the parameter $\alpha>0.2$, the value of the tensor-to-scalar ratio 
$r\rightarrow 0$.

%%%%%%%%%%%%%%%% ADDITION  RAMON -  END  %%%%%%%%%%%%%%%%%%
%%%%%%%%%%%%%%%%%%%%%%%%%%%%%%%%%%%%%%%%%%%%%%%%%%%%%%%%%%%%%%%%
%% LAST ADDITION RAMON - 2
Besides, in particular for the values
$\cP_S\simeq 2.4\times 10^{-9}$ and  $N_{*}=60$ (recall the subscript $*$ indicating 
the epoch where the cosmological scale exits the horizon)  we
obtained for the parameter $\chi_2$ from Eq.(\ref{chi2})  that
$\chi_2\simeq 74\times 10^{-3}$, which corresponds to the value of
$\alpha=0.2$, and $\chi_2\simeq 58\times 10^{-6}$, which
corresponds to the parameter $\alpha=10^{-20}$. In this form the
constraint for $\chi_2$ is given by $58\times
10^{-6}\lesssim\chi_2\lesssim 74\times 10^{-3}$. Here, we have
used  the same values of $b$, $f_1$, $f_2$ and $\epsilon$ from Fig.3.

Numerically, from Eq.\rf{ns-2} we find a
constraint for the parameter $f_1$ given by
$f_1\simeq7.74\times10^{-8}$ for the values $n_s=0.96$ and the
number $N_{*}=60$, which corresponds to the value of $\alpha=0.2$,
and $f_1\simeq 3.58\times10^{-8}$, which corresponds to
$\alpha=10^{-20}$. In this way, the range of the parameter $f_1$
is $3.58\times10^{-8}\lesssim f_1\lesssim 7.74\times10^{-8}$. As
before, we have considered the same values of $b$, $f_2$ and
$\epsilon$ from Fig.3.

%%%%%%%%%%%%%%%%%%%%%%%%%%%%%%%%%%%%%%%%%%%%%%%%%%%%%%%%%%%%%
%%%%%%%%%%%%%%%%%%%%%%%%%%%%%%%%%%%%%%%%%%%%%%%%%%%%%%%%%%%%%
\section{Non-Singular Emergent Universe Solution}
\label{emergent}

We will now show that under appropriate restrictions on the parameters there
exist an epoch preceding the inflationary phase. Namely, we derive an explicit 
cosmological solution of the Einstein-frame system
with effective scalar field Lagrangian \rf{L-eff-final}-\rf{U-eff} describing 
a non-singular ``emergent universe'' \ct{emergent-univ} 
when the scalar field evolves on the first flat region for large negative 
$\vp$ \rf{U-minus}. For previous studies of ``emergent universe'' scenarios within 
the context of the less general modified-measure gravity-matter theories
with one non-Riemannian and one standard Riemannian integration measures, see
Ref.\ct{TMT-recent-1-a}-\ct{TMT-recent-1-c}.

Emergent universe is defined through the standard 
Friedman-Lemaitre-Robertson-Walker space-time metric \rf{FLRW} as a solution
of \rf{friedman-eqs} subject to the condition on the Hubble parameter $H$:
\be
H=0 \quad \to \quad a(t) = a_0 = {\rm const} \,,\;\; \rho + 3p =0 \quad ,\quad 
\frac{K}{a_0^2} = \frac{1}{6}\rho ~(= {\rm const}) \; ,
\lab{emergent-cond}
\ee
with $\rho$ and $p$ as in \rf{rho-def}-\rf{p-def}:

The emergent universe condition \rf{emergent-cond} implies that the $\vp$-velocity
$\vpdot \equiv \vpdot_0$ is time-independent and satisfies the bi-quadratic 
algebraic equation:
\be
\frac{3}{2} B_{(-)}\vpdot_0^4 + 2 A_{(-)}\vpdot_0^2 - 2 U_{(-)} = 0
\lab{vpdot-eq}
\ee
(with notations as in \rf{U-minus}-\rf{A-B-minus}), whose solution read:
\be
\vpdot_0^2 = - \frac{2}{3B_{(-)}} \Bigl\lb A_{(-)} \mp
\sqrt{A_{(-)}^2 + 3 B_{(-)}U_{(-)}}\Bigr\rb \; .
\lab{vpdot-sol}
\ee
Let us note that according to \rf{A-B-minus} $B_{(-)} < 0$ for a wide range of the
parameters, in particular, within the allowed interval of stability (see
\rf{param-constr} below). 
We also observe that under the emergent universe condition \rf{emergent-cond},
and since now $\vpdot$ is time-independent, the $\vp$-equations of motion
\rf{vp-eqs-full} are identically satisfied.

To analyze stability of the present emergent universe solution:
\be
a_0^2 = \frac{6K}{\rho_0} \quad ,\quad
\rho_0 = \h A_{(-)}\vpdot_0^2 + \frac{3}{4} B_{(-)}\vpdot_0^4 + U_{(-)} \; ,
\lab{emergent-univ}
\ee
with $\vpdot_0^2$ as in \rf{vpdot-sol}, we perturb Friedman Eqs.\rf{friedman-eqs}
and the expressions for $\rho,\, p$ \rf{rho-def}-\rf{p-def} w.r.t.  
$a(t) = a_0 + \d a (t)$ and $\vpdot (t) = \vpdot_0 + \d \vpdot (t)$, but
keep the effective potential on the flat region $U_{\rm eff} = U_{(-)}$:
\br
\frac{\d \addot}{a_0} + \frac{1}{12} (\d\rho + 3 \d p) \quad ,\quad
\d \rho = - \frac{2\rho_0}{a_0} \d a \; ,
\lab{delta-friedman} \\
\d\rho = \(A_{(-)}\vpdot_0 + 3 B_{(-)}\vpdot_0^3\) \d\vpdot =
- \frac{2\rho_0}{a_0} \d a \;\; ,\;\;
\d p = \(A_{(-)}\vpdot_0 + B_{(-)}\vpdot_0^3\) \d\vpdot \; .
\lab{delta-rho-p}
\er
From the first Eq.\rf{delta-rho-p} expressing $\d\vpdot$ as function of $\d a$ and
substituting into the first Eq.\rf{delta-friedman} we get a harmonic
oscillator type equation for $\d a$:
\be
\d \addot + \om^2 \d a = 0 \quad ,\quad
\om^2 \equiv \frac{2}{3}\rho_0\,\frac{\pm\sqrt{A_{(-)}^2 + 3B_{(-)}U_{(-)}}}{A \mp
2\sqrt{A_{(-)}^2 + 3 B_{(-)}U_{(-)}}} \; ,
\lab{stability-eq}
\ee
where:
\be
\rho_0 \equiv  
\h\vpdot_0^2 \bigl\lb A_{(-)} + 2\sqrt{A_{(-)}^2 + 3B_{(-)}U_{(-)}}\bigr\rb \; ,
\lab{rho-0}
\ee
with $\vpdot_0^2$ from \rf{vpdot-sol}. Thus, for existence and stability of
the emergent universe solution we have to choose the upper signs in 
\rf{vpdot-sol}, \rf{stability-eq} and we need the conditions:
\be
A_{(-)}^2 + 3B_{(-)}U_{(-)} > 0 \quad ,\quad
A_{(-)} - 2\sqrt{A_{(-)}^2 + 3B_{(-)}U_{(-)}}\bigr\rb > 0 \; .
\lab{stability-cond}
\ee
The latter yield the following constraint on the coupling parameters:
\be
% {\rm max} \Bigl\{-2\,,\, -8\bigl(1+3\eps\frac{f_1^2}{f_2}\bigr)
% \Bigl\lb 1 - \sqrt{1 - \frac{1}{4\bigl(1+3\eps\frac{f_1^2}{f_2}\bigr)}}\Bigr\rb
% \Bigr\} < b\frac{f_1}{f_2} < -1  \; ,
{\rm max} \Bigl\{-2\,,\, -8\bigl(1+3\eps f_1^2/f_2\bigr)
\Bigl\lb 1 - \sqrt{1 - \frac{1}{4\bigl(1+3\eps f_1^2/f_2\bigr)}}\Bigr\rb\Bigr\}
< b\frac{f_1}{f_2} < -1  \; ,
\lab{param-constr}
\ee
in particular, implying that $b<0$. The latter means that both terms in the 
original matter Lagrangian $L^{(2)}$ \rf{L-2} appearing multiplied by the 
second non-Riemannian integration measure density $\Phi_2$ \rf{Phi-1-2} must be
taken with ``wrong'' signs in order to have a consistent physical Einstein-frame 
theory \rf{L-eff-final}-\rf{B-def} possessing a non-singular emergent universe
solution.

For $\eps > 0$, since the ratio $\frac{f_1^2}{f_2}$ proportional to the
height of the first flat region of the effective scalar potential,
\textsl{i.e.}, the vacuum energy density in the early universe, must be
large (cf. \rf{early-vs-late}), we find that the lower end of the interval in 
\rf{param-constr} is very close to the upper end, \textsl{i.e.}, 
$b\frac{f_1}{f_2} \simeq -1$.

%%%%%%%%%%%%%%%% ADDITION 12 - BEGIN  %%%%%%%%%%%%%%%%
From Eqs.\rf{vpdot-sol}-\rf{emergent-univ} we obtain an inequality satisfied by the
initial energy density $\rho_0$ in the emergent universe: 
$U_{(-)} < \rho_0 < 2U_{(-)}$, which together with the estimate of the order
of magnitude for $U_{(-)}$ \rf{U-minus-magnitude} implies order of magnitude for  
$a_0^2 \sim 10^{-8} K M_{Pl}^{-2}$, where $K$ is the Gaussian curvature of the 
spacial section. 
%%%%%%%%%%%%%%%% ADDITION 12 - END  %%%%%%%%%%%%%%%%

For a recent semiclassical analysis of quantum (in)stability of oscillating
emergent universes we refer to \ct{mithani-vilenkin-1,mithani-vilenkin-2}.

%%%%%%%%%%%%%%%%%%%%%%%%%%%%%%%%%%%%%%%%%%%%%%%%%%%%%%%%%%%%%
%%%%%%%%%%%%%%%%%%%%%%%%%%%%%%%%%%%%%%%%%%%%%%%%%%%%%%%%%%%%%
\section{Evolution of the Universe to Its Present Slowly Accelerating State}
\label{evolution}

As a first approach to an unified  analysis of all stages of the
cosmological scenario developed here (emergent universe, transition
from emergent universe to inflation, slow-roll regime, \textsl{etc}.) we
write the set of dynamical equations \rf{friedman-eqs} and \rf{vp-eqs-full} 
as an autonomous system of three dimensions by following the scheme developed in 
Ref.\ct{TMT-recent-1-a}
%%%%%%%%%%%%%%%%%  ADD szydlowski - BEGIN  %%%%%%%%%%%%%%%%%
(for a recent systematic exposition of the methods of dynamical systems' evolution
in the context of cosmology, see \ct{szydlowski}).
%%%%%%%%%%%%%%%%%  ADD szydlowski - BEGIN  %%%%%%%%%%%%%%%%%
We obtain:
%In order to study this point in more details let us
\begin{eqnarray}
\dot{H} &=& - H^2 + \frac{1}{12}\Bigg(2A(\vp)\,x^2 + \frac{3}{2}B(\vp)\,x^4 -
2U_{eff}\Bigg)\;, \label{SD1}\\
\nonumber \\
\dot{x} &=& -\frac{3H\,x\,(A(\vp) + B(\vp)\,x^2) +
\frac{1}{2}A'\,x^2 +  \frac{3}{4}B'\,x^4 + U_{eff}'}{A(\vp) +
3B(\vp)\,x^2}\;,\label{SD2}\\
\nonumber \\
\dot{\vp} &=& x\,,\label{SD3}
\end{eqnarray}
where we have defined $x = \dot{\vp}$. We are considering $\dot{\vp}>0$
because we are interested in the cases where the field $\vp$ moves from
$-\infty$ to positives values, following the emergent universe scheme.

During the emergent universe regime the scalar field evolves on the first flat
region \rf{U-minus} of the effective potential corresponding to large negative $\vp$. 
% which correspond to consider the limit $\alpha\,\vp \rightarrow -\infty$. 
In this case the set of equations (\ref{SD1})-(\ref{SD3}) could be written
as an autonomous system of two dimensions with respect to $H$ and $x$ as follows:
\begin{eqnarray}
\dot{H} &=& - H^2 + \frac{1}{12}\Bigg(2A_{(-)}x^2 +
\frac{3}{2}B_{(-)}\,x^4 -
2U_{(-)}\Bigg)\;, \label{SDC1}\\
\nonumber \\
\dot{x} &=& -\frac{3H\,x\,(A_{(-)} + B_{(-)}x^2)}{A_{(-)} +
3B_{(-)}x^2} \; .\label{SDC2}
\end{eqnarray}

For $x>0$ the above system has six critical points.
In order to study the nature of these critical points we linearize
the set of equations (\ref{SDC1})-(\ref{SDC2}) near the critical
points. From the study of the eigenvalues of the system and by
taking into account the constraints on the pertinent parameters 
\rf{stability-cond}-\rf{param-constr} discussed in the previous Section 5
% discussed in \textbf{Sect.~4} Eqs.~\textbf{(55, 56)}, 
we find that two of these critical points correspond to the emergent universe 
solution, where $H_0 =0$:
\br
x_0^2 &=& \frac{8 b\,f_1 \chi_2 + 16 f_2 \chi_2 \mp \sqrt{(-8 b f_1 \chi_2
-16 f_2 \chi_2)^2 - 16 f_1^2 \Bigg(3 b^2 \chi_2^2- 12 b f_1 \epsilon
\chi_2^2 - 12 f_2 \epsilon  \chi_2^2 \Bigg)}}{2 \Bigg(3 b^2 \chi_2^2 - 12
b f_1 \epsilon \chi_2^2 -12 f_2 \epsilon \chi_2 ^2\Bigg)},
\nonu \\
\phantom{aa} &\phantom{a}& \phantom{aa} 
\lab{SDE1} \\
H_0 &=& 0 \,.
\nonu
\er
In this case, the upper sign in Eq.\rf{SDE1} corresponds to a
center critical point and the lower sign is an unstable saddle
point. The stable emergent universe solution % used in the Emergent Universe scheme
obtained in Section 5 is the stable center critical point.

Also, we have the following critical points which are (upper sign)
attractor and a (lower sign) focus:
\br
x^2_0 &=& \frac{2 (b f_1 + 2 f_2)}{\left(b^2 - 4 b f_1 \epsilon - 4
f_2 \epsilon \right) \chi_2}\,, \\
\nonumber \\
H_0 &=& \pm\frac{\sqrt{\frac{2 f_1^2}{\left(f_2+ f_1^2 \epsilon
\right) \chi_2} - \frac{4 (b f_1 + 2 f_2)}{\left(b^2 - 4 b f_1
\epsilon - 4 f_2 \epsilon \right) \chi_2} - \frac{2 b f_1 (b f_1 +
2f_2)}{\left(f_2 + f_1^2 \epsilon \right) \left(b^2 - 4 b f_1
\epsilon - 4 f_2 \epsilon \right) \chi_2} + \frac{4 f_1^2 (b f_1 + 2
f_2) \epsilon}{\left(f_2+ f_1^2 \epsilon \right) \left(b^2-4 b f_1
\epsilon - 4 f_2 \epsilon \right) \chi_2 }}}{4 \sqrt{3}}.
\nonu \\
\phantom{aa} &\phantom{a}& \phantom{aa} 
\lab{SDE2}
\er
These critical points are similar to the kinetic vacuum state
discussed in Ref.\ct{TMT-recent-1-a}, but in the present case these
critical points only exists in the limit % $\alpha\,
$\vp \rightarrow -\infty$.

%%%%%%%%%%%%%%%%%% Fig 2 %%%%%%%%%%%%%%%%%%%%%%%
\begin{figure}
\centering
\includegraphics[width=7cm]{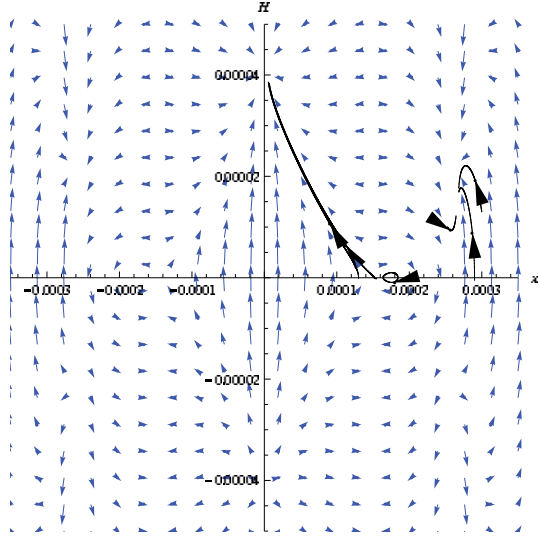}
\caption{Plot showing part of the direction field of the system
Eqs.(\ref{SDC1})-(\ref{SDC2}) and six numerical solutions.
\label{SDF1}}
\end{figure}
%%%%%%%%%%%%%%%%%%%%%%%%%%%%%%%%%%%%%%%%%%%%%%%

On the other hand, we have the standard slow-roll de Sitter critical
point:
\begin{equation}\label{SDE3}
x_0^2=0,\;\;\;\;\; H_0= \pm\frac{f_1}{2 \sqrt{6} \sqrt{f_2 + f_1^2
\epsilon} \sqrt{\chi_2 }},
\end{equation}
where the upper sign is an attractor and the lower sign is a
focus in Eq.(\ref{SDE3}). This is the standard slow-roll de Sitter
attractor.

% In figure \ref{SDF1} 
In Fig.4 it is shown a phase portrait for six numerical
solution to Eqs. (\ref{SDC1})-(\ref{SDC2}), where we have taken
$f_1 = 2 \times 10^{-8}, f_2= 1 \times 10^{-8}, \epsilon = 1, \alpha = 1,
% \chi_2 = 1, M_1 = 4 \times 10^{-64}, M_2 =4$, and $b = - 0.52$. 
\chi_2 = 1, M_1 = 4 \times 10^{-60}, M_2 =4$, and $b = - 0.52$. 
Also, in this figure we have included the
direction field of the system in order to have a visual picture of what a
general solution look like. In Fig.4 the six critical
points described above are depicted. One of this points is the center equilibrium
point ($H = 0, x = 0.00017$), the saddle point ($H = 0, x = 0.00025$), 
the point ($H = 0.000015, x = 0.00026$) is a future
attractor and ($H = -0.000015, x = 0.00026$) is a past attractor.
The other equilibrium points are ($H = 0.000040, x = 0$) and ($H =
-0.000040, x = 0$) which are a future attractor point and a past
attractor point, respectively.

As we have mentioned above, the slow-roll de Sitter critical point is an
attractor, then, it is plausible that some of the solutions near the
center critical point, when the effective potential begins to be
nonconstant, start to move away from the center critical point and
finish at the slow-roll de Sitter critical point; see Fig.5. % figure \ref{SDF2}. 
During this short period, which occurs before the
slow-roll period, the Hubble parameter satisfies $\dot{H} >0$. This
period is called ``super-inflation'' and has been studied in the
context of emergent universe scenario in Ref.\ct{labrana}. After this
short period the system arrives at the slow-roll regime discussed in Section 3 above.

%%%%%%%%%%%%%%%%%% Fig 1 %%%%%%%%%%%%%%%%%%%%%%%
\begin{figure}
\centering
\includegraphics[width=7cm]{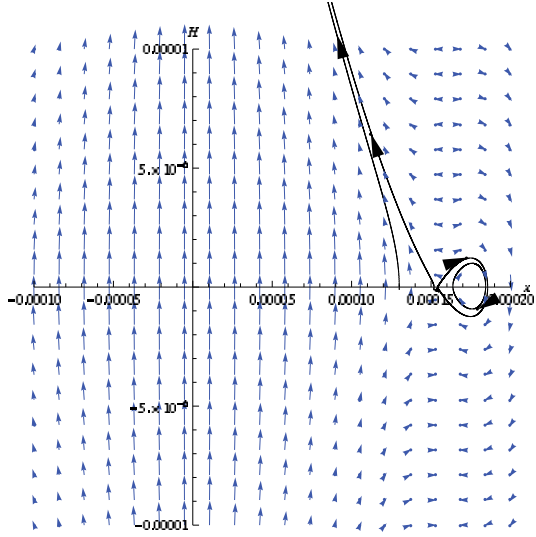}
\caption{Plot showing part of the direction field of the system and
four numerical solutions, near the center critical point.
\label{SDF2}}
\end{figure}
%%%%%%%%%%%%%%%%%%%%%%%%%%%%%%%%%%%%%%%%%%%%%%%

%%%%%%%%%%%%%%%%%% Fig 3 %%%%%%%%%%%%%%%%%%%%%%%
\begin{figure}
\centering
\includegraphics[width=7cm]{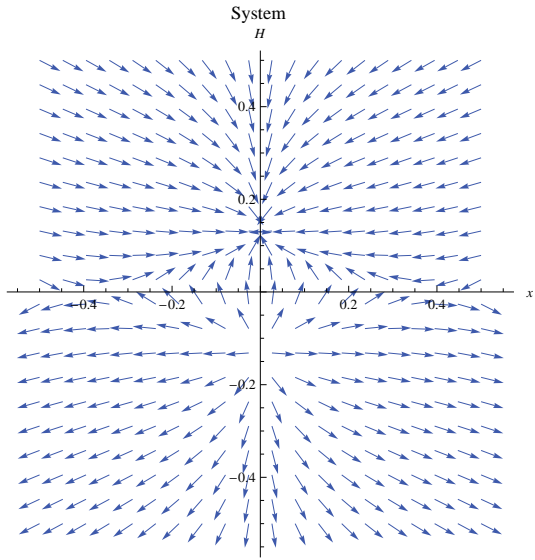}
\caption{Plot showing part of the direction field of the system
Eqs.(\ref{Pinf1})-(\ref{Pinf2}). \label{SDF3}}
\end{figure}
%%%%%%%%%%%%%%%%%%%%%%%%%%%%%%%%%%%%%%%%%%%%%%%

On the other hand, during the present slowly accelerating phase of
the universe, the scalar field evolves on the second flat region of
the effective potential \rf{U-plus} corresponding to large positive $\vp$.
% which correspond to consider the limit$\alpha\,\vp \rightarrow \infty$. 
In this case the set of Eqs.(\ref{SD1})-(\ref{SD3}) can be 
written as an autonomous system of two dimensions as follows:
\begin{eqnarray}
\dot{H} &=& \frac{M_1^2 \left(1-24 H^2 \epsilon \,\chi_2 \right)- M_2
\chi_2 \left(24 H^2+4 x^2+3 x^4 \epsilon \, \chi_2 \right)}{24 \left(M_2
+ M_1^2\,\epsilon \right) \chi_2}\;, \label{Pinf1}\\
\nonumber \\
\dot{x} &=& -\frac{3 H x \left(1+x^2 \epsilon \, \chi_2 \right)}{1+3
x^2 \epsilon \chi_2 } \;.\label{Pinf2}
\end{eqnarray}
The system has two critical points:
\begin{eqnarray}
H_0 &=& \pm\frac{M_1}{2 \sqrt{6} \sqrt{M_2+ M_1^2 \epsilon}
\sqrt{\chi_2 }}\,, \\ x_0 &=& 0 \,.
\end{eqnarray}
These critical points are an attractor and a focus respectively.
% In figure \ref{SDF3} 
In Fig.6 it is shown the qualitative shape of the  direction field of the system
Eqs.(\ref{Pinf1})-(\ref{Pinf2}) near the critical points. 
In particular, we have considered the values 
$M_1 = 2$ and $M_2 = 6$ instead of the values considered previously, in 
order to have a more clear view of the critical points in the direction 
field.
% In order to have a more clear
% picture, in this case we have taken $f_1 = 2 \times 10^{-8}, f_2=
% 1 \times 10^{-8}, \epsilon = 1, \alpha = 1, \chi_2 = 1, M_1 = 2 , %\times 10^{-64}, 
% M_2 = 6$, and $b = - 0.52$.
%
In this figure there are the two critical points described above. One of
this point is the future attractor ($H = 0.129, x = 0$), and the
other is the past attractor point ($H = -0.129, x = 0$).

%%%%%%%%%%%%%%%%%%%%%%%%%%%%%%%%%%%%%%%%%%%%%%%%%%%%%%%%%%%%%
%%%%%%%%%%%%%%%%%%%%%%%%%%%%%%%%%%%%%%%%%%%%%%%%%%%%%%%%%%%%%
\section{Discussion}
\label{discuss}
% =====================
% Two flat regions of effective scalar potential -- Unification of Inflation and 
% Dark Energy from Spontaneous Breaking of global Weyl-Scale Invariance
% ====================

In the present paper we have constructed a new kind of gravity-matter theory
defined in terms of two different non-Riemannian volume-forms (generally
covariant integration measure densities) on the space-time manifold, where
the Einstein-Hilbert term $R$, its square $R^2$, the kinetic and the
potential terms in the pertinent cosmological scalar field (a ``dilaton'')
% , as well as the standard Maxwell Lagrangian and a ``square-root'' Maxwell term of an
% associated (Abelian) gauge field 
couple to each of the non-Riemannian
integration measures in a manifestly globally Weyl-scale invariant form.
The principal results are as follows:

\begin{itemize}
\item
Dynamical spontaneous symmetry breaking of the global Weyl-scale invariance.
\item
In the physical Einstein frame we obtain an effective scalar field potential
with {\em two flat regions} -- one corresponding to the early universe evolution and 
a second one for the present slowly accelerating phase of the universe.
\item
The flat region of the  effective scalar potential appropriate for describing 
the early universe allows for the existence of a {\em non-singular ``emergent''} type
beginning of the universe' evolution. This ``emergent'' phase is followed,
%%%%%%%%%%%%%%%%%%%% SUPERINFLATION - BEGIN %%%%%%%%%%%%%%%
via a short period of ``super-inflation'', 
%%%%%%%%%%%%%%%%%%%% SUPERINFLATION - END %%%%%%%%%%%%%%%
by the inflationary phase, which in turn is followed by a period, where the
scalar field drops from its high energy density state to the present slowly
accelerating phase of the universe.
%%%%%%%%%%%%%%% ADDITION  PLANCK DATA - BEGIN  %%%%%%%%%%%%%%%%%
\item
For a reasonable choice of the parameters the resulting ratio of tensor-to-scalar
perturbations conforms to the data of Planck Collaboration.
%%%%%%%%%%%%%%% ADDITION  PLANCK DATA - END  %%%%%%%%%%%%%%%%%
\end{itemize}

The flatness of the effective scalar potential in the high energy density
region makes the slow rolling inflation regime possible.

The presence of the emergent universe' phase preceding the inflationary
phase has observable consequences for the low CMB multipoles as has been
recently shown in Ref.\ct{labrana}. 
% Therefore, a full analysis of the CMB
% results in the context of the present model should involve not only the
% classical ``slow-roll'' formalism, but also the ``super-inflation'' one, which
% describes the transition from the emergent universe to the inflationary phase.
% For a systematic unified analysis of all stages of the cosmological scenario
% developed here (emergent universe, transition from emergent universe to inflation --
% period of ``super-inflation'', ``slow roll'' regime, \textsl{etc.}) it will be
% instructive to employ the methods of dynamical systems' evolution \ct{szydlowski}.

%%%%%%%%%%%%%%%%%%%%%%%%%%%%%%%%%%%%%%%%%%%%%%%%%%%%%%%%%%%%%%%%
%% LAST ADDITION RAMON - 2
Table 1 summarizes the constraints on the parameters in
the different phases in the context of the present model. Let us note that
although we don't have separate constraints on $M_1$ and $M_2$, nevertheless on
Section 3 we made the natural choice to identify them with the two
fundamental scales $M_{EW}$ and $M_{Pl}$, which then yielded the correct
order of magnitude \rf{U-plus-magnitude} of the present epoche's dark energy
dominated vacuum energy density. Similarly, although the inflationary phase
only determines the scale of the ratio $f_1^2/f_2$ \rf{U-minus-magnitude}
we made the natural choice for these parameters setting $f_1 \sim f_2$ to be 
of the same order of magnitude since both originally appear as coupling 
constants in front of two scalar field potential terms of the same type.
For the last parameter $\eps$ we have found the restriction 
$|\eps| \frac{M_1^2}{M_2} \ll 1$ (second inequality in \rf{early-vs-late-2}).

%%%%%%%%%%%%%%% TABLE - BEGIN %%%%%%%%%%%%%
\begin{table}
\begin{tabular}
[c]{||c||c||c||}\hline Phase & Constraint from & Constraint on
\\\hline\hline Dark energy dominated
 & vacuum energy density Eq.(34) &
\begin{tabular}
[c]{c}%
$\frac{M_1^2}{M_2}\simeq 10^{-120}M^4_{Pl}$\\
%$M_2\simeq M^4_{Pl}$\\
%
\end{tabular}
%&
\\
\hline\hline
Inflation %\\
(using also emergent) &
\begin{tabular}
[c]{c}%
%\\
Eq.\rf{U-minus-magnitude}\\
$\cP_S\simeq 2.4\times 10^{-9}$ and $N_*=60$\\
$n_s=0.96$ and $N_*=60$\\
consistency relation $n_s=n_s(r)$%
\end{tabular}
&
\begin{tabular}
[c]{c}%
\\
$\frac{f_1^2}{f_2}\sim10^{-8}M^4_{Pl}$\\ \\
$58\times10^{-6}\lesssim\chi_2\lesssim 74\times 10^{-3}$\\ \\
$3.6\times10^{-8}\lesssim f_1\lesssim 7.7\times 10^{-8}$\\ \\
$0\lesssim\alpha\lesssim 0.2$\\ \\%
\end{tabular}
\\\hline\hline
Non-singular emergent &
\begin{tabular}
[c]{c}%
upper end of the interval in Eq. (69)\\
\end{tabular}
&
\begin{tabular}
[c]{c}%
$b\frac{f_1}{f_2}\simeq -1$\\
\end{tabular}
%&
\\\hline%\hline
\end{tabular}
\caption{Results for the constraints on the parameters in our model.
}
\end{table}
%%%%%%%%%%%%%%% TABLE - END %%%%%%%%%%%%%

%%%%%%%%%%%%%% OSCILLATIONS ADD - BEGIN  %%%%%%%%%%%%%%%%%%%%%%%
We conclude with some comments of qualitative nature.
The oscillations  of the scalar field $\vp$ are an important  part for the 
standard mechanism of reheating of the Universe \cite{reh1}. However, when
the integration constant $M_1 <0$ our effective scalar potential \rf{U-eff}
does not have a minimum (cf. Fig.1) so that the scalar field $\vp$ does not 
oscillate and, therefore, the standard reheating does work. 
In the literature these models are known as non-oscillating (NO) models \cite{reh2}. 
An option for the mechanism of reheating in these NO models is the instant
preheating which inserts an interaction between the scalar field driving the
inflationary scenario and another scalar field \cite{reh3}. 
Other mechanism of reheating for the NO models is the insertion  of the 
curvaton field \cite{reh4}. Here the decay of the curvaton  into conventional
matter offers an effective  reheating scenario, and  does not introduce an 
interaction between the inflaton field and  another scalar field \cite{reh5,reh6}. 
In a future work we will study the extension of the present model to include a 
curvaton field according to the basic principles of two non-Riemannian volume forms
on the underlying spacetime and of spontaneous breakdown of global Weyl-scale 
invariance. 
%%%%%%%%%%%%%% OSCILLATIONS ADD - END  %%%%%%%%%%%%%%%%%%%%%%%

%%%%%%%%%%%%%%  ADD ESCAPE MIN - BEGIN  %%%%%%%%%%%%%%%%%%%%%%%
When the integration constant $M_1 >0$ the effective scalar potential
\rf{U-eff} possesses an absolute minimum $U_{\rm eff} = 0$ at 
$\vp = \vp_{\rm min}$, where $\exp \{-\a\vp_{\rm min}\}= M_1/f_1$ (cf. Fig.2).
%%%%% RAMON CHANGE - BEGIN  %%%%%%
As it is evident from Fig.2, there is an abrupt fall to $U_{\rm eff} = 0$ where 
% presumably 
particle creation will take place % is obtained from rapidly varying $\vp (t)$. 
when we consider the extended theory enlarged with a curvaton as mentioned above.
%%%%% RAMON CHANGE - END %%%%%%
The scalar field falls down with very high kinetic energy into the region of 
$U_{\rm eff} \simeq 0$, the kinetic energy before the fall-down being certainly 
vastly higher than the value of $U_{\rm eff} \simeq U_{(+)}$ \rf{U-plus-magnitude} 
in the second flat region to the right. 
So $\vp (t)$ ``climbs'' the latter very low barrier and continues to 
evolve in the $\vp \to +\infty$ direction. Thus, on the second flat region we have
a slow rolling scalar field which produces approximately the dark energy equation of state
$\rho \simeq - p$, with very small $\rho = U_{(+)}$ \rf{U-plus-magnitude}
explaining the present day dark energy phase.
In a future work we plan to study in more details the evolution of the scalar
field in the vicinity of, and its escape out of, the global minimum $U_{\rm eff} = 0$.
%%%%%%%%%%%%%%  ADD ESCAPE MIN - END  %%%%%%%%%%%%%%%%%%%%%%%

%%%%%%%%%%%%%%%%%%%%%%%%%%%%%%%%%%%%%%%%%%%%%%%%%%%%%%%%%%%%%
%%%%%%%%%%%%%%%%%%%%%%%%%%%%%%%%%%%%%%%%%%%%%%%%%%%%%%%%%%%%%
\begin{acknowledgements}
We express our gratitude to Alexander Kaganovich for collaboration at the
initial stage of this work \ct{emergent-belgrade}.
We are thankful to Alexei Starobinsky, Marek Szydlowski, Martin Cederwall
and Lilia Anguelova for instructive discussions. We are indebted to the
referee for his constructive suggestions to improve the present work.
E.G., E.N. and S.P. gratefully acknowledge support of our collaboration through the 
academic exchange agreement between the Ben-Gurion University in Beer-Sheva,
Israel, and the Bulgarian Academy of Sciences. 
R.H. was supported by Comisi\'on Nacional de Ciencias y
Tecnolog\'{\i}a of Chile through FONDECYT Grant 1130628 and DI-PUCV 123.724.
P.L. was supported by Direcci\'on de Investigaci\'on de la
Universidad del B\'{\i}o-B\'{\i}o through grants GI121407/VBC and 141407 3/R.
S.P. has received partial support from European COST action MP-1210.
\end{acknowledgements}
%%%%%%%%%%%%%%%%%%%%%%%%%%%%%%%%%%%%%%%%%%%%%%%%%%%%%%%%%%%%

\end{document}